\documentclass[amsfonts,prl,aps,superscriptaddress]{revtex4}
\usepackage{graphicx}
\usepackage{epsfig}
\usepackage{amsfonts}
\usepackage{amsmath}
\usepackage{amssymb}
\begin{document}
\title{Approximate $\sqrt t$ relaxation in glass-forming liquids} 
\author{Albena I. Nielsen}
\affiliation{DNRF Centre ``Glass and Time,'' IMFUFA, Department of Sciences, Roskilde University, Postbox 260, DK-4000 Roskilde, Denmark} 
\author{Tage Christensen}
\affiliation{DNRF Centre ``Glass and Time,'' IMFUFA, Department of Sciences, Roskilde University, Postbox 260, DK-4000 Roskilde, Denmark} 
\author{Bo Jakobsen}
\affiliation{DNRF Centre ``Glass and Time,'' IMFUFA, Department of Sciences, Roskilde University, Postbox 260, DK-4000 Roskilde, Denmark} 
\author{Kristine Niss}
\affiliation{DNRF Centre ``Glass and Time,'' IMFUFA, Department of Sciences, Roskilde University, Postbox 260, DK-4000 Roskilde, Denmark}
\author{Niels Boye Olsen}
\affiliation{DNRF Centre ``Glass and Time,'' IMFUFA, Department of Sciences, Roskilde University, Postbox 260, DK-4000 Roskilde, Denmark} 
\author{Ranko Richert}
\affiliation{Department of Chemistry and Biochemistry, Arizona State University, Tempe, Arizona 85287-1604, USA}
\author{Jeppe C. Dyre}
\affiliation{DNRF Centre ``Glass and Time,'' IMFUFA, Department of Sciences, Roskilde University, Postbox 260, DK-4000 Roskilde, Denmark} 
\date{\today}
\newcommand{\tv}{\tilde\varepsilon}
\newcommand{\ft}{\tilde f}
\newcommand{\ve}{\varepsilon}
\newcommand{\am}{\alpha_{\rm min}}
\newcommand{\fm}{f_{\rm max}}
\newcommand{\id}{I_{\Delta E}}
\begin{abstract}
We present data for the dielectric relaxation of 43 glass-forming organic liquids, showing  that the primary (alpha) relaxation is often close to $\sqrt t$ relaxation. The better an inverse power-law description of the high-frequency loss applies, the more accurately is $\sqrt t$ relaxation obeyed. These findings suggest that $\sqrt t$ relaxation is generic to the alpha process, once a common view, but since long believed to be incorrect. Only liquids with very large dielectric losses deviate from this picture by having consistently narrower loss peaks. As a further challenge to the prevailing opinion, we find that liquids with accurate $\sqrt t$ relaxation cover a wide range of fragilities.
\end{abstract}

\maketitle

Glass may be regarded as the fourth state of conventional matter, isotropic as the liquid state, but solid as the crystalline state. With the notable exception of helium, any liquid may be turned into glass by cooling it fast enough to avoid crystallization \cite{gut95,edi96,ang00,don01,dyr06}. At the glass transition all molecular motion beyond vibrations essentially come to a standstill. Since glass inherits the liquid's molecular structure, a deeper understanding of the glassy state may come from understanding the highly viscous liquid phase preceding glass formation. This paper presents a compilation of data for glass-forming organic liquids probed by dielectric relaxation, i.e., by studying their response to an external periodic electric field. The data show that the current understanding of viscous-liquid dynamics is incomplete, and that these dynamics are simpler than presently believed.

Physical systems usually relax following perturbations forced upon them. The simplest form of relaxation is an exponential decay towards equilibrium which is, however, surprisingly seldom observed. Another simple case is the so-called $\sqrt t$ relaxation where the relaxation function $\Phi(t)$ at short times decays as $\Phi(0)-\Phi(t)\propto\sqrt t$. This is observed in systems as diverse as free Rouse dynamics of polymer chains \cite{doi86}, molecular nanomagnets \cite{wer99,wer06}, and turbulent transport, e.g., in astrophysics \cite{bak04}. For random walks, the equivalent of $\sqrt t$ relaxation is referred to as single-file diffusion which is observed, e.g., in ion channels through biological membranes, diffusion in zeolites, and charge-carrier migration in one-dimensional polymers \cite{mon02}. Below we present data showing prevalence of $\sqrt t$ relaxation in glass-forming liquids. The data were taken on organic liquids in the extremely viscous state approaching the glass transition; here the relaxation time in some cases is larger than 1 second -- which may be contrasted to the relaxation time of ambient water that is around 1 picosecond. This is the first proof of a general prevalence of $\sqrt t$ relaxation in highly viscous liquids, thus going significantly beyond the conjecture from 2001 that if a liquid obeys accurate time-temperaure superposition, it exhibits $\sqrt t$ relaxation \cite{ols01}.

The dominant and slowest relaxation process of a glass-forming liquid is the so-called alpha process. The alpha process defines the liquid relaxation time, an important quantity because the glass transition takes place when the relaxation time significantly exceeds the inverse relative cooling rate. The most accurate data for the alpha process are obtained by dielectric relaxation measurements that directly probe molecular rotation in liquids where the molecules have a permanent dipole moment. With modern frequency analyzers precise data may be obtained over more than 9 decades of time/frequency \cite{kre02}, and large amounts of dielectric data for glass-forming liquids are now available \cite{kud99,rol05}. 

We have measured the dielectric loss $\varepsilon''$ as a function of frequency for a number of organic glass formers and supplemented these by data from the Bayreuth and Augsburg groups, giving altogether 43 liquids measured slightly above the glass transition temperature \cite{suppl}. The only liquids excluded are monoalcohols \cite{note_1}, which are known to exhibit a very strong Debye loss peak with a relaxation time that is not coupled to the calorimetric glass transition \cite{hut07}. In order to avoid bias, data were selected prior to their analysis. A model-free data analysis was performed, i.e., without fitting data to any of the standard functions (stretched exponential, Havriliak-Nagami, Cole-Cole, Cole-Davidson, etc). This is an unusual procedure because very accurate data are required in order to obtain, e.g., reliable slopes by numerical differentiation. The data analysis was automated as far as possible via Matlab programs. We did not distinguish between liquids with and without clearly visible Johari-Goldstein secondary (beta) processes, but included all available data of sufficient quality to complete the analysis. No attempts were made to subtract possible contributions from beta processes. Details of the measurements are given in Ref. \cite{suppl}.

\begin{figure}\begin{center}
\includegraphics[width=8cm]{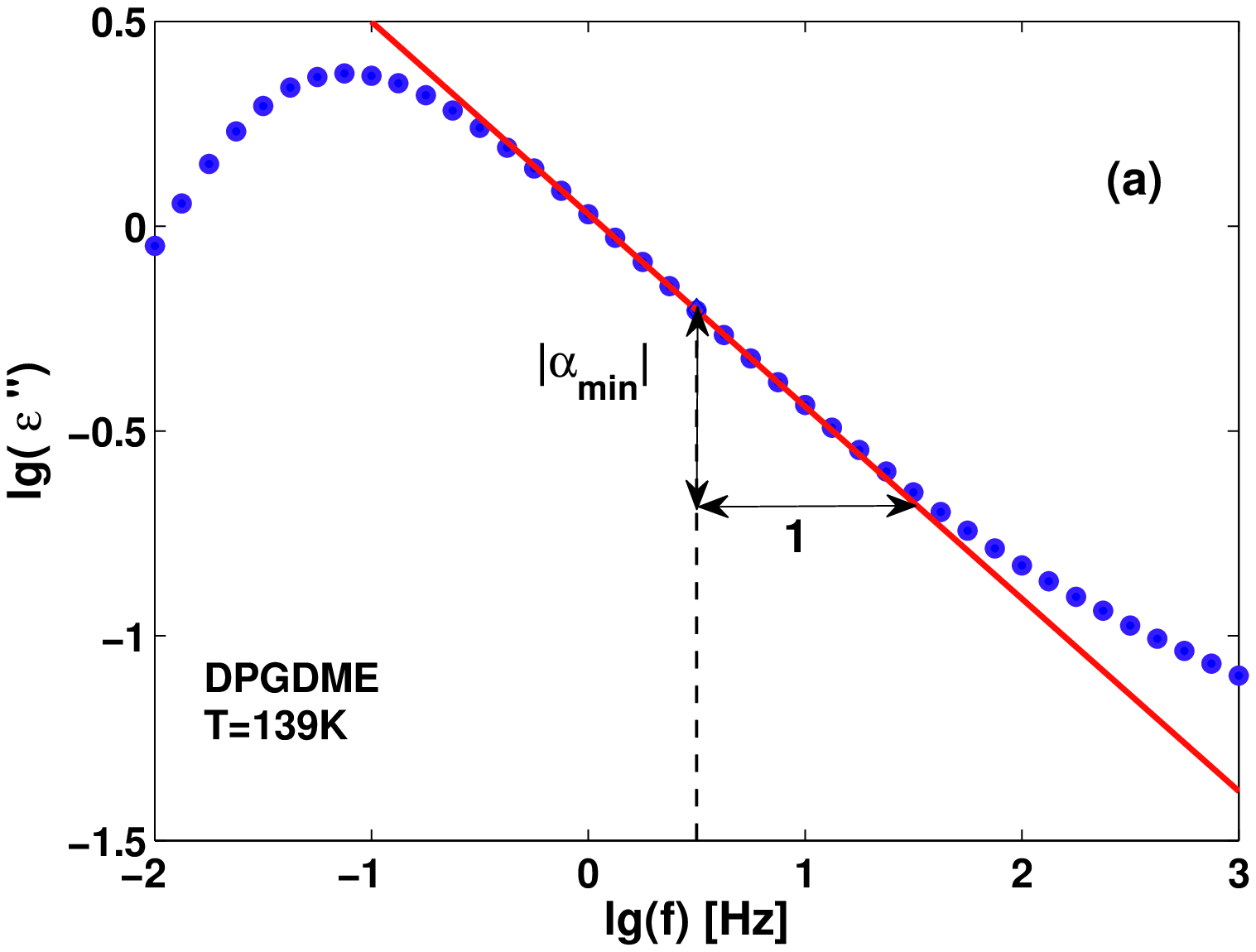}
\includegraphics[width=8cm]{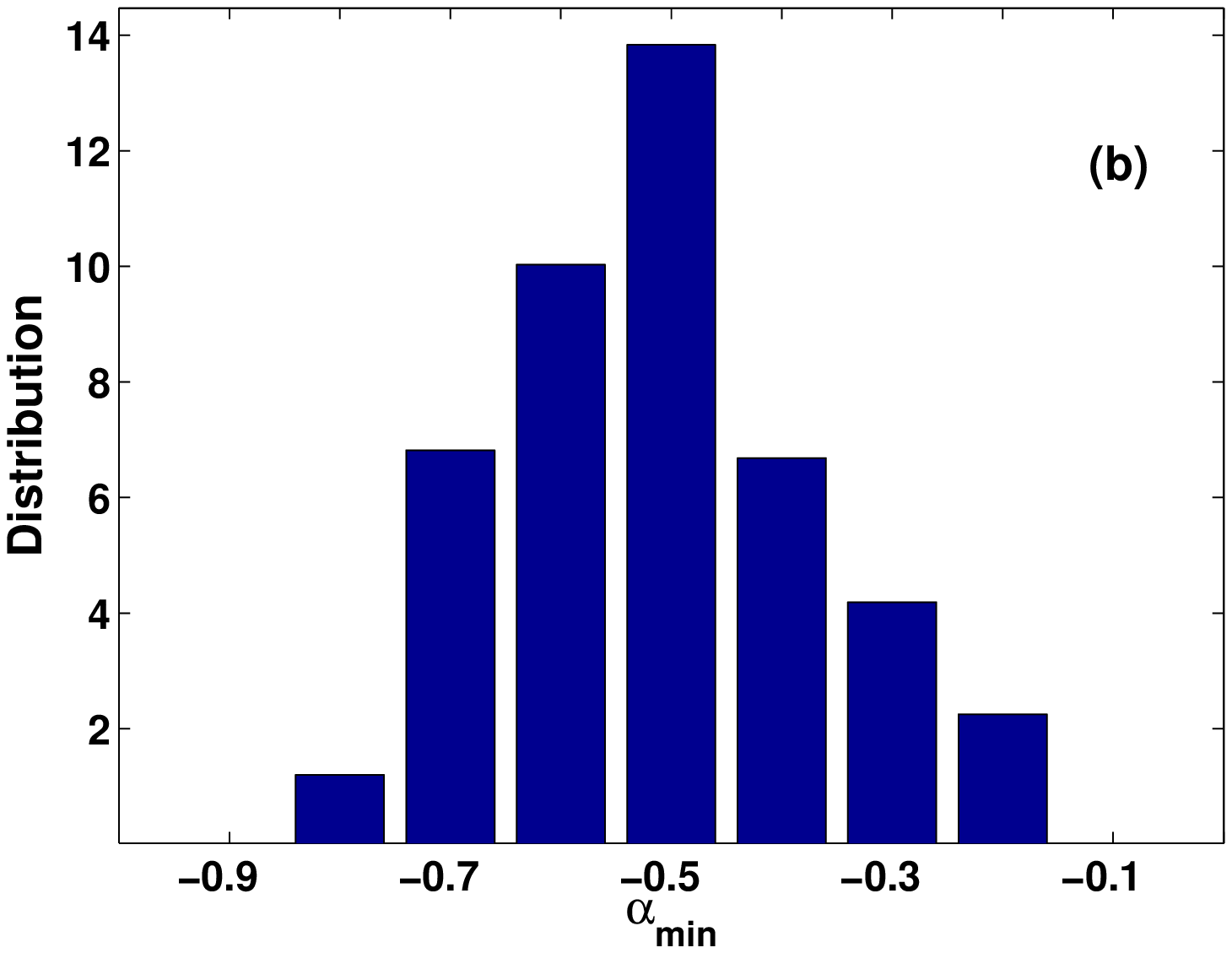}
\caption{(a): Log-log plot of the dielectric loss $\varepsilon''$ as a function for frequency for Dipropylene Glycol Dimethyl Ether at $139$ K. The red line is the tangent at the inflection point that gives a good approximation to data over two decades of frequency. The vertical dashed line indicates the position of the inflection point, i.e., of the minimal slope $\am$ (in this example $\am=-0.46$). (b): Histogram of the minimum slope distribution for 43 organic glass-forming liquids. The figure shows one column for $-0.85<\am<-0.75$, one for $-0.75<\am<-0.65$, etc. Since the number of temperatures investigated varies from liquid to liquid, each minimum slope observation is weighted by a factor $1/N$ where $N$ is the number of temperatures for the given liquid. In this way all liquids contribute equally to the histogram. Note that a liquid may contribute to more than one column in this figure since $\am$ may vary with temperature.}
\end{center}\end{figure}

The low-frequency (long-time) properties of the alpha process are fairly trivial; the vast majority of glass-forming liquids here exhibit what corresponds to a cut-off in the relaxation time distribution function at long times \cite{kre02,kud99,rol05}. Focusing on the short-time (high-frequency) relaxation properties, at each temperature we identified the {\it minimum slope} in the standard log-log plot, $\am\equiv\min\left(d\lg \ve''/d\lg f\right)<0$, where $f$ is frequency and $\lg$ is the base-10 logarithm. This identifies the inflection point above the loss-peak frequency. As illustrated in Fig. 1(a) the number $\am$ gives the best approximate inverse power-law description of the loss decay above the peak: $\ve''(f)\propto f^{-|\am|}$ applies to a good approximation over a significant frequency range. Only data with a well-defined minimum slope or a clear slope plateau were used, and only if there is so little noise in the point-by-point numerical differentiation that data allow a precise determination of $\am$. For the data thus selected Fig. 1(b) shows a histogram of the minimum-slope distribution. The above-mentioned limitations as well as the different temperature ranges and intervals used by various groups imply that the number of data sets per liquid varies widely (from 2 to 26). To compensate for this and give equal weight to each liquid in Fig. 1(b), if $N$ data sets were included in the analysis for a given liquid, each minimal-slope observation was weighted by a factor $1/N$ for this liquid.

\begin{figure}\begin{center}
\includegraphics[width=8cm]{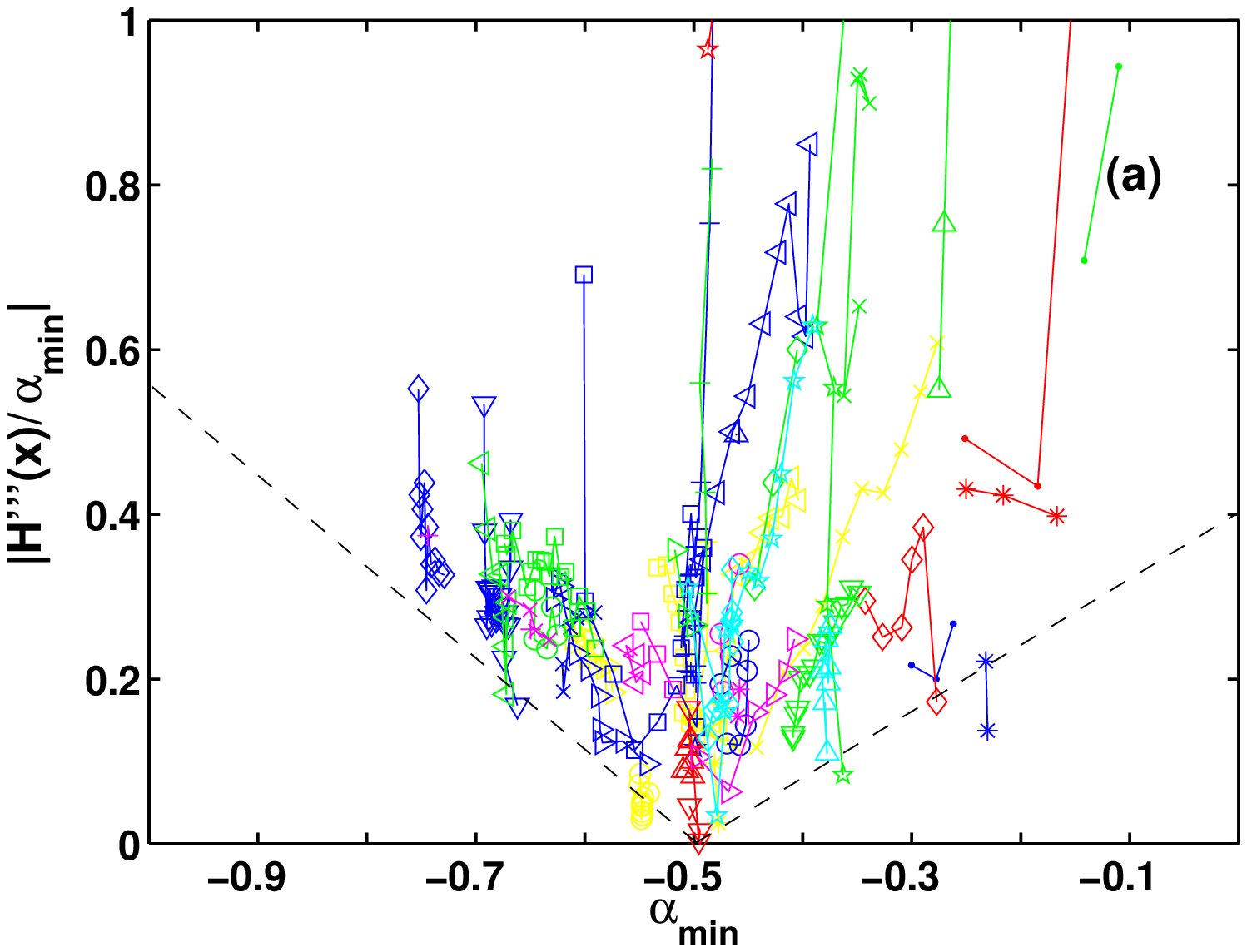}
\includegraphics[width=8cm]{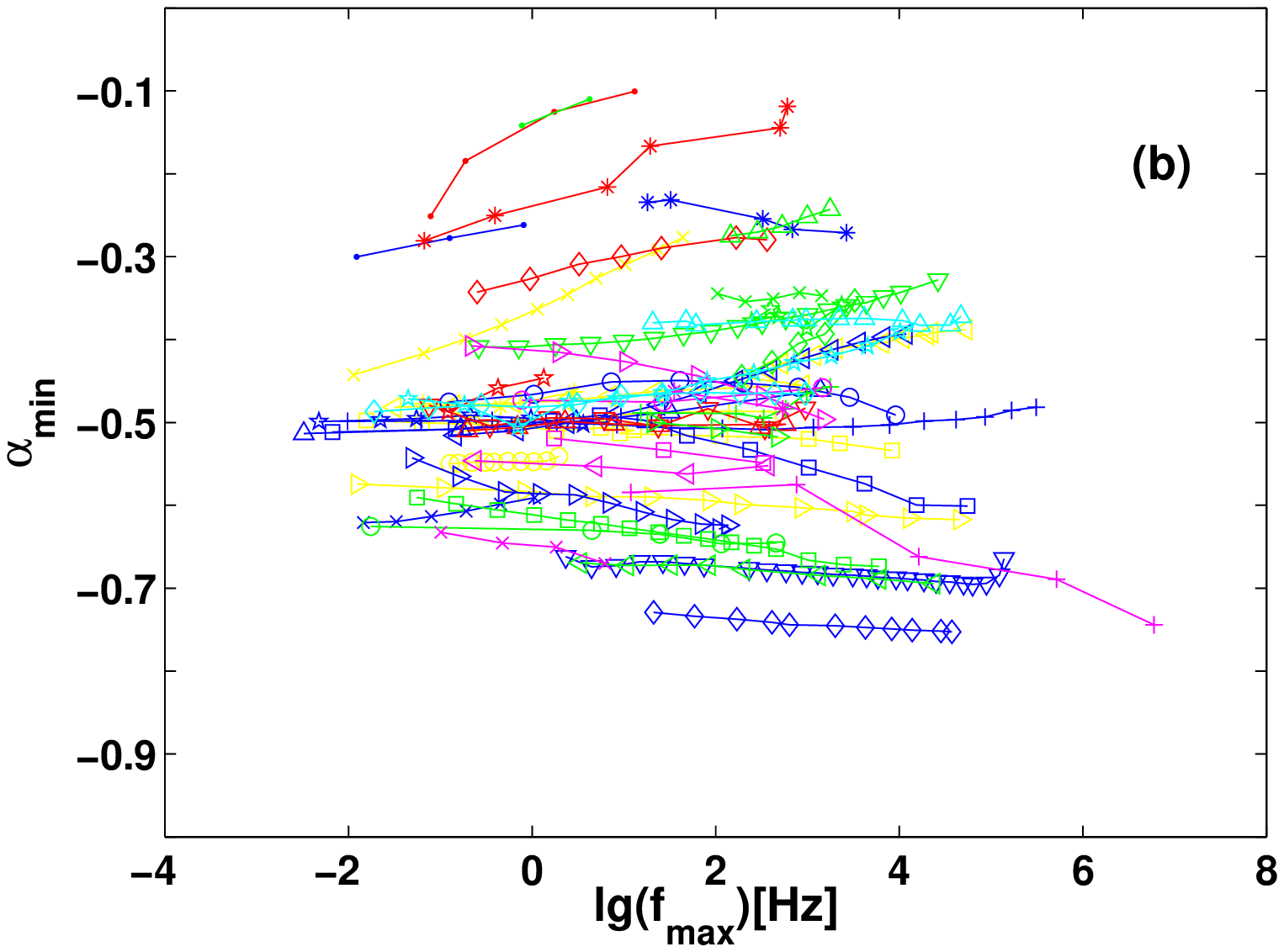}
\caption{(a): Third-order relative to first-order derivative, $|H^{(3)}(x)/\am)|$, at the frequency of minimum slope for all data sets where $H(x)\equiv\lg \ve''(x)$ ($x=\lg  f)$. At the frequency of minimum slope the second-order derivative is zero; thus by Taylor's formula the smaller the third-order derivative is relative to the first-order derivative $\am=H'(x)$, the better an inverse power law description of the high-frequency loss applies. The dashed lines are drawn as guides to the eye. (b): The minimum slope $\am$ plotted as a function of temperature quantified by the position of the loss peak frequency $\fm$ for all 43 liquids.}
\end{center}\end{figure}

According to the fluctuation-dissipation theorem an $f^{-1/2}$ high-frequency dielectric loss implies that the dipole time-autocorrelation function $\Phi_P(t)$ for times shorter than the alpha-relaxation time obeys $\Phi_P(0)-\Phi_P(t)\propto\sqrt t$. Thus liquids with $\am\simeq -1/2$ exhibit approximate $\sqrt t$ relaxation of the dipole autocorrelation function. An obvious question is whether the observed prevalence of minimum slopes around $-1/2$ is a coincidence. If $\am=-1/2$ were significant, one would expect that the closer the minimum slope is to $-1/2$, the better an inverse power-law description applies. This is investigated in Fig. 2(a) that plots the third-order derivative relative to the first-order derivative, $|H^{(3)}(x_0)/\am|$, where $H(x)=\lg \ve''(x)$, $x=\lg  f$, and $x_0$ is the log frequency at the point of minimum slope. The idea is that, since the second-order derivative is zero at the frequency of minimum slope, by Taylor's formula the smaller $|H^{(3)}(x_0)/\am|$ is, the larger is the frequency range where the slope is almost constant. Figure 2(a) shows that the better an inverse power law describes the loss, the closer is $\am$ to $-1/2$.

\begin{figure}\begin{center}
\includegraphics[width=8cm]{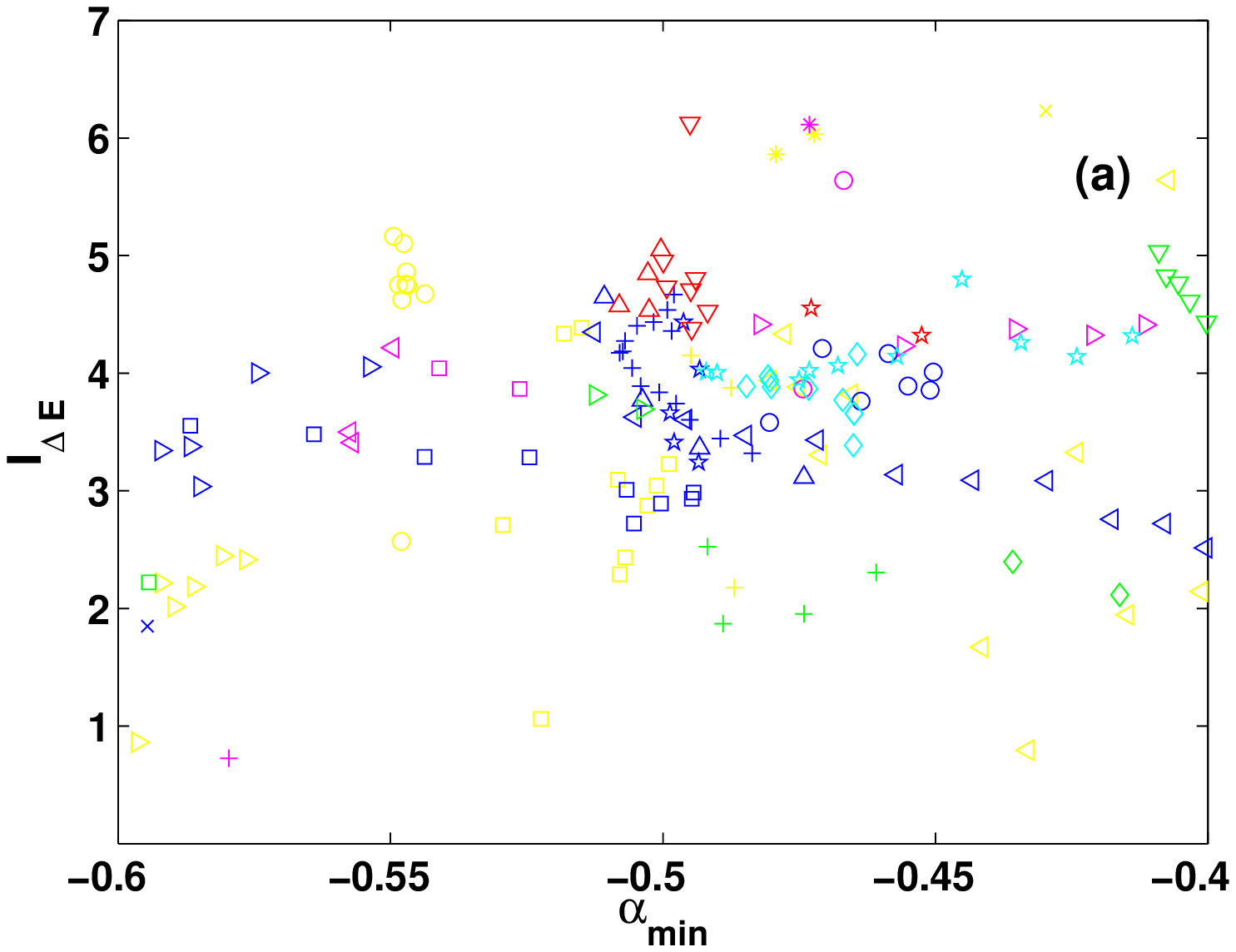}
\includegraphics[width=8cm]{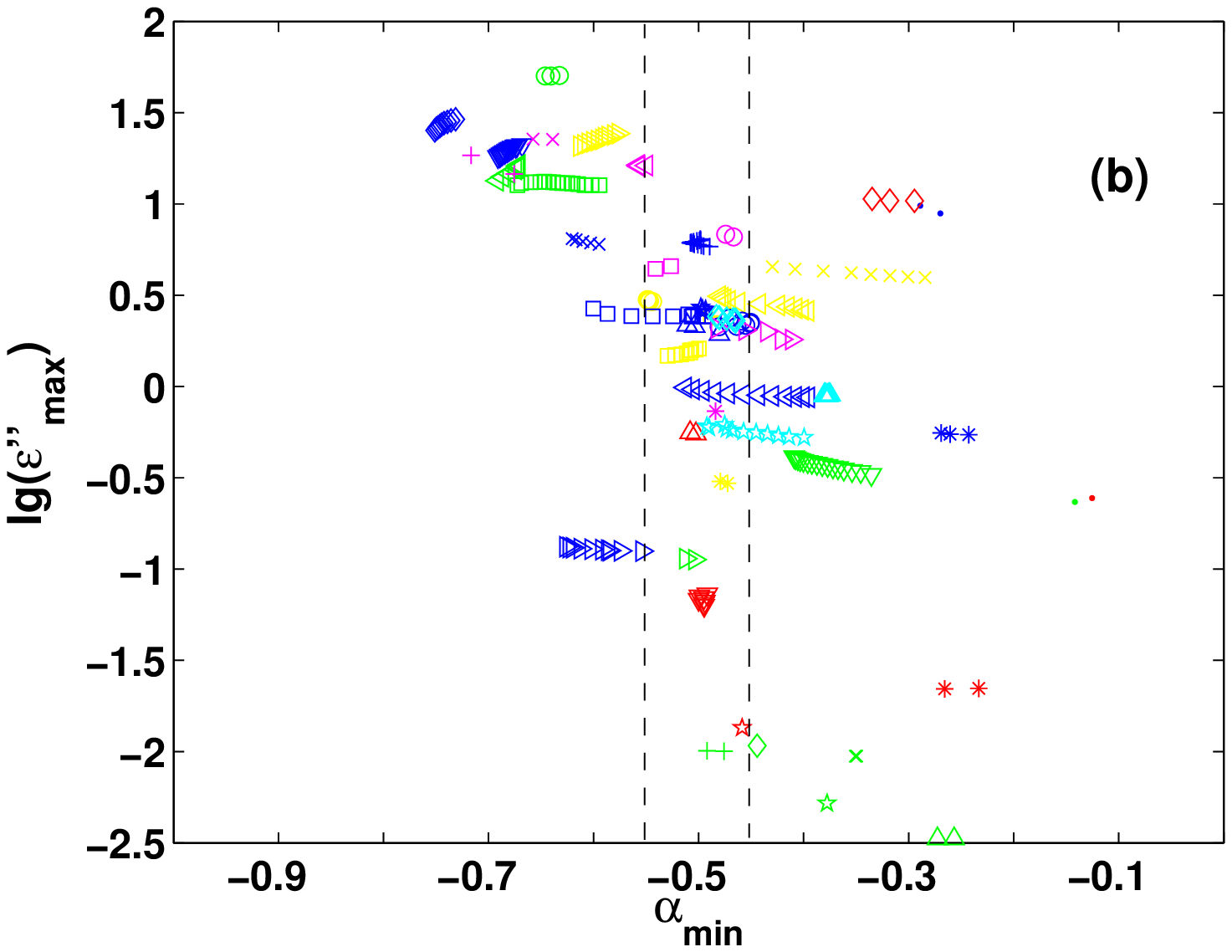}
\caption{(a): The ``activation energy temperature index'' $\id\equiv|d\ln\Delta E/d\ln T|$ versus $\am$ for all data sets with $-0.6<\am<-0.4$, i.e., those  obeying approximate $\sqrt t$ relaxation. The quantity $\id$ is a measure of the degree of deviation from Arrhenius temperature dependence of the loss-peak frequency similar to Angell's fragility $m$, but defined at all temperatures. A broad range of fragilities is represented among liquids obeying $\sqrt t$ relaxation, thus right at $\am=0.5$ the temperature index varies by a factor of $2.5$. In terms of fragility this takes one from $50$ to $125$ which is practically the entire span of fragilities. (b): Maximum dielectric loss versus $\am$ for all data sets. The liquids between the two dashed lines marking the interval $-0.55<\am<-0.45$ have dielectric losses varying by more than a factor of 1,000. Very large-loss liquids have minimum slopes that numerically are significantly larger than $1/2$; these liquids do not exhibit approximate $\sqrt t$ relaxation.}
\end{center}\end{figure}

Figure 2(b) plots the minimum slope as a function of temperature represented by the loss-peak frequency $\fm$. Although not without exceptions, the overall tendency is that the minimum slopes converge slowly to $-1/2$ as temperature is lowered \cite{ols01}. From Fig. 2 we conclude that the prevalence of $\sqrt t$ relaxation of Fig. 1(b) is hardly coincidental, but reveals a generic property of viscous liquid dynamics. Figure 3 investigates other physical quantities that $\am$ might correlates with. According to the generally accepted picture, relaxation properties are determined by how much the temperature dependence of a liquid's relaxation time deviates from the Arrhenius equation. More precisely, if the dipole time-autocorrelation function is fitted by a stretched exponential, $\exp[-(t/\tau)^\beta]$, according to the conventional wisdom the larger the liquid fragility is, the lower is $\beta$ \cite{boh93}. Since a stretched exponential implies a high-frequency power-law loss varying with frequency as $f^{-\beta}$, one expects liquids exhibiting $\sqrt t$ relaxation to have only a narrow range of fragilities. This is tested in Fig. 3(a) where the activation energy temperature index $\id(T)\equiv|d\ln\Delta E/d\ln T|$ \cite{sch98,dyr04} is plotted for all data sets with $-0.6<\am<-0.4$, where the activation energy was identified by writing $\fm(T)=f_0\exp(-\Delta E(T)/k_BT)$ with $f_0=10^{-13}$ s. $\id(T)$ provides a measure of how non-Arrhenius a liquids is; the connection to the Angell's fragility $m$ is provided \cite{dyr04} by $m=16(\id(T_g)+1)$. Figure 3(a) shows that liquids with $\sqrt t$ relaxation exhibit a wide range of fragilities.

In Fig. 3(b) we plot the minimum slope versus maximum loss $\varepsilon''(\fm)$, the latter quantity reflecting the overall dielectric loss strength. Liquids covering more than 4 decades of dielectric strengths are represented in the data set. Liquids with very large dielectric strength show minimum slopes that are consistently larger than $1/2$ numerically. These liquids present an exception to $\sqrt t$ relaxation. The liquids with $|\am|>0.65$ all have Kirkwood correlation factors that are significantly larger than unity, reflecting strong correlations between the motions of different dipoles. Higher Kirkwood correlation factors promote longer range orientational and dynamical correlations, leading to spatial averaging of what might otherwise still be $\am=-1/2$ behavior (in fact, for $g\rightarrow\infty$ one expects an approach to Debye relaxation reflecting an increasingly large degree of cooperativity).

In conclusion, these data suggest that, with the exception of very large-loss liquids, $\sqrt t$ relaxation is generic to the alpha process of glass-forming liquids; $\sqrt t$ relaxation is a better approximation, the better an inverse power law describes the high-frequency loss (Fig. 2(a)) and the lower the temperature is (Fig. 2(b)). If approximate $\sqrt t$ relaxation were generic for glass-forming liquids, however, how does one explain the fact that $40\%$ of the data do not obey $-0.6<\am<-0.4$? The most likely explanation is that deviations are caused by interference from Johari-Goldstein beta (secondary) relaxation processes with no distinct maximum. Although for many years it was believed that secondary processes are found only in the kHz-MHz frequency range, it is now generally recognized that theses processes in some cases take place at much lower frequencies \cite{ols98,sch00}. If one or more secondary processes appear in the Hz range, it is practically impossible to separate alpha and beta processes \cite{note_2}. -- Recently $\sqrt t$ relaxation was reported for the mechanical relaxation of liquids forming bulk metallic glasses \cite{wen04}, as well as for the shear-mechanical relaxation of glass-forming organic liquids \cite{jak05}. It should be emphasized, however, that mechanical relaxation measurements on highly viscous liquids are much less accurate than dielectric measurements.

Although our findings go against presently prevailing views of viscous liquid dynamics, it is not a new idea that a high-frequency loss varying as $f^{-1/2}$ is generic to highly viscous liquids. In the 1960's and beginning of the 1970's this was, in fact, commonly noted (see, e.g.,  Ref. \cite{wil75}). Several theories were proposed to explain it, including Glarum's defect diffusion model \cite{gla60,dor70}, the ``inhomogeneous media'' model of Isakovich and Chaban \cite{isa66}, the Barlow-Erginsav-Lamb (BEL) model postulating a mechanical equivalent of a simple electrical circuit \cite{bar67}, and the Montrose-Litovitz model invoking diffusion and relaxation of some unspecified order \cite{mon70}. The idea gradually fell out of favor to be replaced by the presently prevailing view that relaxation functions are basically determined by the fragility \cite{boh93}. Our data show a prevalence of $\sqrt t$ relaxation independent of fragility, however. If $\sqrt t$ relaxation is eventually confirmed as generic for the alpha process (excluding high-loss liquids), the dynamics of glass-forming liquids is simpler than generally believed. That presents an obvious challenge to theory.

\acknowledgments 
The authors wish to thank Ernst R{\"o}ssler, Alois Loidl, and Peter Lunkenheimer for kindly providing their data. This work was supported by a grant from the Danish National Research Foundation (DNRF) for funding the centre for viscous liquid dynamics ``Glass and Time.''


\begin{thebibliography}{99}
\bibitem{note_1} Polymers were excluded from the analysis because the polymer glass transition is not a liquid-glass transition.
\end{thebibliography}
\end{document}